\documentclass[twocolumn,showpacs,preprintnumbers,amsmath,amssymb,nofootinbib]{revtex4-1}
\usepackage{graphicx}
\usepackage{dcolumn}
\usepackage{float}
\usepackage{url}
\usepackage{siunitx}
\usepackage{amsfonts}

\newcommand{\abs}[1]{\left|#1\right|}

\begin{document}

\title{Synthesis of optical spring potentials in optomechanical systems}
\author{Harry J.\ Slatyer, Giovanni Guccione, Young-Wook Cho, Ben C.\ Buchler and Ping Koy Lam}\email{Ping.Lam@anu.edu.au}
\affiliation{Centre for Quantum Computation and Communication Technology, Department of Quantum Science, Research School of Physics and Engineering, The Australian National University, Canberra ACT 2601, Australia}
\date{\today}
\begin{abstract}
	We propose a method to tailor the potential experienced by a moveable end mirror in a cavity optomechanical system by specifying the spectral properties of the input field. We show that by engineering the power spectral density of the cavity input field a desired force function can be approximated, with the accuracy of the approximation limited only by the linewidth of the cavity. The very general technique presented here could have applications in many kinds of optomechanical systems, particularly those used for sensing and metrology. We demonstrate the method by applying it to improve the sensitivity of a particular gravity measurement.
\end{abstract}

\maketitle

\section{Introduction}
\label{sec: Introduction}

In the last few years there have been a great number of experiments and theoretical proposals demonstrating the power and flexibility of optomechanical systems. This includes metrology applications, for example force~\cite{Gavartin:2012:NatNano,Harris:2013:PhysRevLett,Hosseini:2014:NatComm} and mass~\cite{Li:2012:ApplPhysLett,Liu:2013:OptExp} measurement, and fundamental tests of quantum mechanics, such as the observation of semiclassical gravity~\cite{Yang:2013:PhysRevLett} and spontaneous wave-function collapse~\cite{Nimmrichter:2014:PhysRevLett}. A particularly promising concept is the design of optically levitated mechanical oscillators~\cite{Barker:2010:PhysRevLett,Kiesel:2013:PNAS,Neukirch:2014:ContempPhys,Shvedov:2014:NatPhot}, since the near-complete isolation from the environment can result in extremely low mechanical dissipation rates.

A Gaussian beam incident on a micron-scale dielectric particle tends to trap the particle in the region of highest optical intensity due to the difference in force experienced by opposite sides of the particle refracting light of different intensity~\cite{Ashkin:1970:PhysRevLett,Ashkin:2000:IEEEJSelTopQuantumElectron}. This can be used for levitation \cite{Ashkin:1971:ApplPhysLett} and is also the basis of optical tweezers, which have become an important tool for micro- and nano-scale manipulation \cite{Moffitt:2008:AnnuRevBiochem,Marago:2013:NatNano}. Recent work has also explored the optomechanical properties of particles in such traps~\cite{Li:2011:NatPhys,Millen:2014:NatNano}. The optical potential experienced by the trapped particle can be tuned by shaping the transverse mode of the levitating lasers~\cite{Dholakia:2011:NatPhot}, or using an optical cavity~\cite{Chang:2010:PNAS,Kiesel:2013:PNAS} to modify the longitudinal mode of the light. Ultimately, however, the performance of this system in cooling and sensing applications will be limited by the scattering of light that is inherent to their operation.

The scattering could be eliminated if the levitated object were to be a mirror in a high-finesse optical cavity~\cite{Singh:2010:PhysRevLett,Guccione:2013:PhysRevLett}. A further feature of this system is that the optical spring effect~\cite{Braginsky:1997:PhysLettA,Sheard:2004:PhysRevA,Corbitt:2007:PhysRevLett} enables precise control of the optical potential experienced by the mirror and therefore tuneability of the mechanical properties of its oscillation \cite{Schediwy:2008:PhysRevA}. Although precise control is possible, the range of possible optical spring parameters is fixed by the finesse of the cavity. A high-finesse cavity leads to very stiff spring constants, which may not always be the desired outcome. For example, in a sensing application where the position of the levitated mirror is used to measure some force, large mechanical response is required to maximise the signal. Ideally, therefore, one desires an optical spring of lower stiffness, while still using a high-finesse cavity to maintain optimum interferometric sensitivity of the position readout.

In this paper we propose a general method to allow for greater flexibility and performance in cavity-based optomechanical systems by showing how customized force functions (and hence also customized potentials) may be approximated by specifying the spectral properties of the input field. The concept of engineering optical force functions by injecting polychromatic light has already been suggested~\cite{Rakich:2009:OptExp}, but we extend this idea to show exactly which polychromatic light sources are necessary to generate particular force functions. We use the standard Lorentzian peak of an optical cavity as a building block to create more elaborate force profiles that can be tailored to meet specific requirements of an experiment. Since the ideal required power spectral densities may be difficult or impossible to generate in practice, we show that approximation of these by appropriate frequency combs is possible.

In section~\ref{sec: Force from a single frequency} we review the physics behind the optomechanical force from a conventional single input. Before extending these principles to multiple inputs, we consider the implications of interference terms in section~\ref{sec: The interaction of multiple optical springs}. An approximation method for the desired force function is then presented in sections~\ref{sec: Approximation of an arbitrary force function} and~\ref{sec: Practical considerations}. As a concrete example of an application of this technique, in section~\ref{sec: An application} we consider a gravitational sensor based on an optomechanical levitation system (a simplified version of that proposed in \cite{Guccione:2013:PhysRevLett}) and show how the sensitivity can be enhanced with no additional input power or reduction in stability.

\section{Force from a single frequency}
\label{sec: Force from a single frequency}

In this section we derive the force experienced by the end mirror in an optical cavity as a function of its position and the laser detuning, which will form the basis of the main analysis.

\begin{figure}[]
	\centerline{\includegraphics[width=\columnwidth]{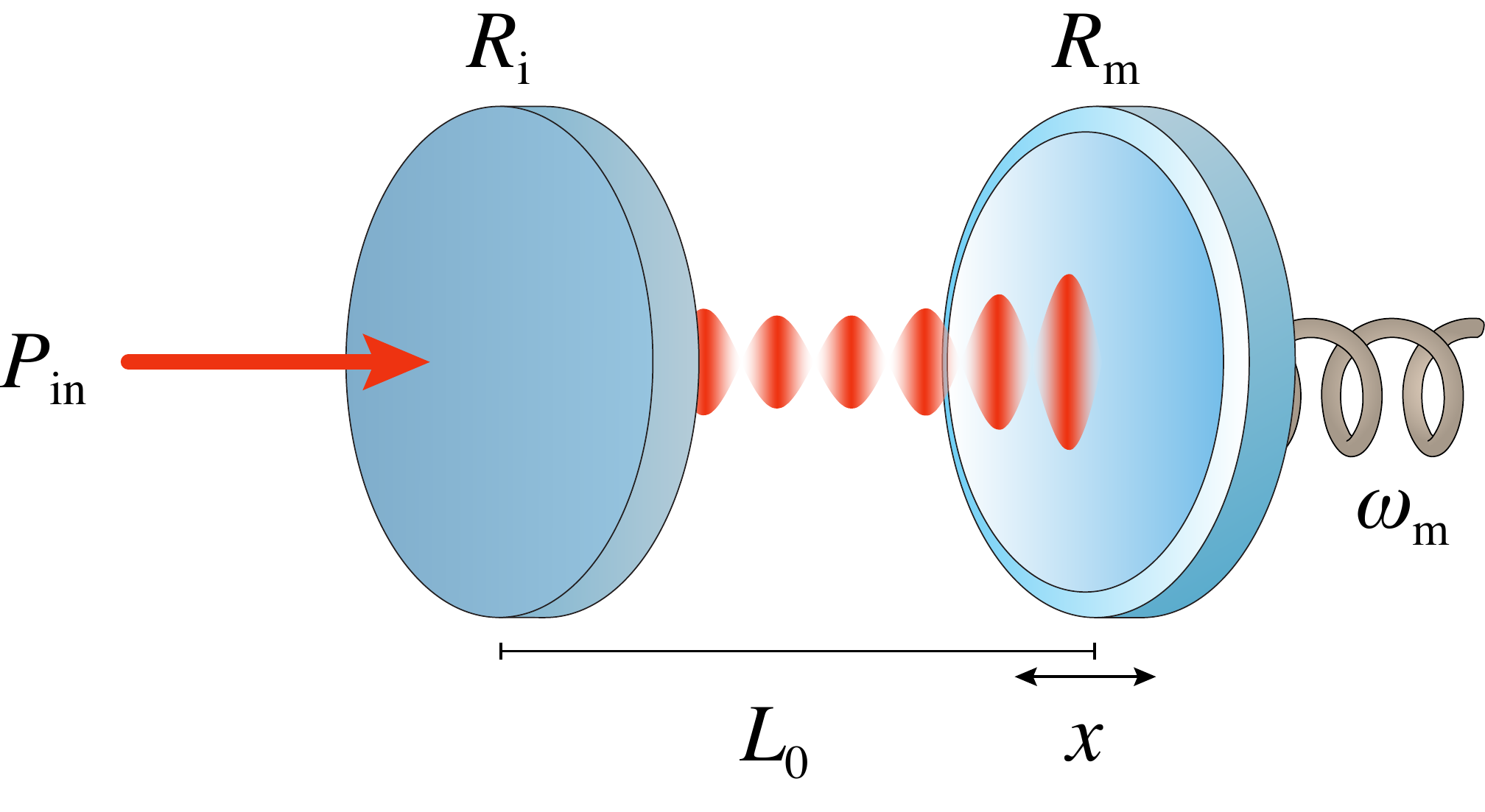}}
	\caption[]{Schematic of the proposed system: an optical cavity with a fixed input mirror of reflectivity $R_\textrm{i}$ and a moveable mirror of reflectivity $R_\textrm{m}$. The total length of the cavity, $L_x=L_0+x$, depends on the position of the moveable mirror. The input may consist of a single-mode or a multi-mode field, as discussed in Section~\ref{sec: The interaction of multiple optical springs}.}
\label{fig: diagram}
\end{figure}

We consider a simple linear cavity with a moveable end mirror such as the one presented in Fig.~\ref{fig: diagram}. Let the reflectivities of the fixed input mirror and moveable end mirror be $R_\textrm{i}$ and $R_\textrm{m}$ respectively, and let $L_x:=L_0+x$ denote the length of the cavity depending on the displacement $x$ of the moveable mirror from its mean position along the optical axis. Without loss of generality assume $L_0$ to be a multiple of $\lambda_0/2$ to achieve resonance when there is no displacement. Also, let the laser frequency be $\omega_\delta:=\omega_0+\delta$, where $\delta$ is the detuning from some central frequency $\omega_0$. The laser wavelength is then given by $\lambda_\delta=2\pi c/\omega_\delta$, where $c$ is the speed of light.

The power $P_x(\delta)$ circulating in the cavity is known to be~\cite{Siegman:1986:USciBooks}, in terms of the input power $P_\textrm{in}$,
\begin{equation}
	P_x(\delta)=\frac{1-R_\textrm{i}}{1+R_\textrm{i}R_\textrm{m}-2\sqrt{R_\textrm{i}R_\textrm{m}}\cos\left(2\pi\frac{L_x}{\lambda_\delta/2}\right)}P_\textrm{in}.	\label{eqn: firstpeq}
\end{equation}
The argument of the cosine term expands to
\begin{equation}
	2\pi\frac{L_x}{\lambda_\delta/2} = 2\pi\frac{L_0}{\lambda_0/2}+2\pi\frac{x}{\lambda_0/2}+2\frac{L_0\delta}{c}+2\frac{x\delta}{c}.
\end{equation}
Since $\frac{L_0}{\lambda_0/2}\in\mathbb{Z}$, the first phase term may be ignored. Next, assuming $\abs{x}\ll\lambda_0/2$ and $\abs{x\delta}\ll L_0\abs{\delta}\ll c$, which is accurate if we consider only one free spectral range, we can approximate the cosine as
$$
	\cos\left(2\pi\frac{x}{\lambda_0/2}+2\frac{L_0\delta}{c}+2\frac{x\delta}{c}\right)\simeq1-\frac{1}{2}\frac{4\pi^2}{\omega_\textrm{FSR}^2}\left(\delta+g_0x\right)^2,
$$
where we have introduced the free spectral range of the cavity $\omega_\textrm{FSR}:=\pi c/L_0$ and the optomechanical coupling $g_0:=2\omega_\textrm{FSR}/\lambda_0$ converting a displacement of the mirror into a cavity frequency detuning. Defining $\delta_x:=\delta+g_0x$, the intra-cavity power can now be expressed as
$$
	P_x(\delta)=\frac{1-R_\textrm{i}}{\left(1-\sqrt{R_\textrm{i}R_\textrm{m}}\right)^2+\sqrt{R_\textrm{i}R_\textrm{m}}\frac{4\pi^2}{\omega_\textrm{FSR}^2}\delta_x^2}P_\textrm{in}.
$$
The linewidth $\gamma$ of the cavity, satisfying $P_0(\gamma/2)=P_0(0)/2$, is given by
$$
	\gamma=\frac{1-\sqrt{R_\textrm{i}R_\textrm{m}}}{\pi\sqrt[4]{R_\textrm{i}R_\textrm{m}}}\omega_\textrm{FSR}=\frac{\omega_\textrm{FSR}}{\mathcal F},
$$
where $\mathcal F$ is the finesse. Substituting into the expression for $P_x(\delta)$ we have
$$
	P_x(\delta)=\frac{1-R_\textrm{i}}{\left(1-\sqrt{R_\textrm{i}R_\textrm{m}}\right)^2}\frac{\gamma^2/4}{\gamma^2/4+\delta_x^2}P_\textrm{in}.
$$
The radiation pressure force $F_x(\delta)$ on the end mirror is therefore
\begin{align}
	F_x(\delta)	&	=\frac{1+R_\textrm{m}}{c}P_x(\delta)															\nonumber\\
				&	\simeq\frac{2P_\textrm{in}}{c}\frac{1-R_\textrm{i}}{\left(1-\sqrt{R_\textrm{i}R_\textrm{m}}\right)^2}\frac{\gamma^2/4}{\gamma^2/4+\delta_x^2},	\label{eqn: F_x definition}
\end{align}
where in the last line we have approximated $1+R_\textrm{m} \approx 2$, assuming the mirror reflectivity $R_\textrm{m}$ to be close to unity as required for a cavity of reasonable finesse.

The optical spring constant is related to the derivative of the force with respect to the position of the mirror,
\begin{align}
\frac{dF_x(\delta)}{dx}	&	=-\frac{2P_\textrm{in}}{c}\frac{1-R_\textrm{i}}{(1-\sqrt{R_\textrm{i}R_\textrm{m}})^2}\frac{2g_0\delta_x\gamma^2/4}{(\gamma^2/4+\delta_x^2)^2}	\nonumber\\
					&	=-\frac{2P_x(\delta)}{c}\frac{2g_0\delta_x}{\gamma^2/4+\delta_x^2}.	\nonumber
\end{align}
Accounting for a delayed response of the system~\cite{Braginsky:2002:PhysLettA,Corbitt:2007:PhysRevLett}, the optical spring at given position and detuning is
\begin{align}
	k_\textrm{os}(\omega)=\frac{2P_x(\delta)}{c}\frac{2g_0\delta_x}{\gamma^2/4+\delta_x^2}\left(1-\frac{\omega^2+i\gamma\omega}{\gamma^2/4+\delta_x^2}\right)^{-1}.	\label{eqn: optical spring}
\end{align}
The complex nature of the optical spring has a double effect on the system: when the sign of the detuning is chosen so that the stiffness, represented by the real part of $k_\textrm{os}$, is positive (blue detuning), the force on the mirror is restoring; however, in the same regime the imaginary part is responsible for the creation of an anti-damping component, and the system may display parametric amplification of the oscillations.

\section{The interaction of multiple optical springs}
\label{sec: The interaction of multiple optical springs}

The force profile described by Eq.~\ref{eqn: F_x definition} is valid only for a single-mode input. Generally speaking, two or more different fields injected into the cavity will interfere and cause a component of the total intra-cavity power to beat. The force experienced by the mirror will be subject to the same beating, and cannot be inferred by the simple sum of the forces experienced from each field separately.

The mechanical response of the system, however, may be less or more receptive to interference depending on the time scale of the beating. A mechanical oscillator has a strong response around the natural mechanical frequency, $\omega_\textrm{m}$, in a bandwidth determined by the damping rate, $\gamma_\textrm{m}$, of the oscillator. If the beat frequency is many multiples of $\gamma_\textrm{m}$ higher than $\omega_\textrm{m}$, then the oscillator perceives only a time-averaged effect of the interference. The resulting steady-state force function in this case corresponds to the sum of the force functions from single-mode inputs (see the appendix for a numerical justification of this claim).

Suppose we wish to create a composite optical spring with two optical fields, but the frequency difference between them leads to beating at \emph{low} frequencies that will drive the mirror. What can be done to eliminate the effects of the beating and allow the addition of the optical springs? Detuning one of the fields by an integer multiple of the free spectral range, $\omega_\textrm{FSR}$, will not alter the circulating power thanks to the periodicity of Eq.~\ref{eqn: firstpeq}. Furthermore, if the applied detuning is large compared to $\omega_\textrm{m}$, the beating will no longer drive the motion of the mirror. As explained above, in this case the force due to individual fields may be simply added. For example, a linear cavity with round-trip length $2L_0=\SI{10}{\centi\metre}$ gives $\omega_\textrm{FSR}=2\pi\times3$~GHz. In this case a shift by a single free spectral range will be much larger than any likely mechanical frequency.

This idea extends easily from the case of two input fields to the case of a frequency comb input. Suppose a spacing $\Delta$ between the modes of the comb. We may identify an integer $N$ such that $N\Delta\gg\omega_\textrm{m}$. Beating between modes separated by $N\Delta$ will not drive the motion of the mirror. For any modes separated by less than $N\Delta$ we can apply the method outlined above and shift the frequency of one mode by a multiple of the free spectral range. The process is then iterated over all the modes in the comb until each free spectral range only carries modes that are spectrally separated by more than $N\Delta$. Formally, we shift the $n$th peak of the comb by $(n\;\text{modulo}\;N)$ multiples of $\omega_\textrm{FSR}$ so that a total of $N$ free spectral ranges are employed, each hosting a number of modes equal to the total number of modes of the comb divided by $N$. This technique ensures that, provided $\omega_\textrm{FSR}$ is much higher than $\omega_\textrm{m}$, every pair of modes beats at a frequency much higher than $\omega_\textrm{m}$, and therefore that interference effects do not drive the mirror. Note that if this assumption on $\omega_\textrm{FSR}$ does not hold then we simply need to shift each peak by a higher multiple of $\omega_\textrm{FSR}$.

Therefore, given any frequency comb input, we can perform frequency shifting so that the sum of the forces due to each individual mode well approximates the averaged effective force experienced by the mirror. In other words, we can simply assume that for \emph{any} frequency comb input field the superposition property holds, with the understanding that any necessary frequency shifting is included implicitly.

Importantly, this method does not extend to the case of a general continuous power spectral density (PSD) input, since in that case there is a continuum of ``peaks'' that must be frequency-shifted. We will see that this is not problematic, since the class of frequency comb input fields is sufficiently broad for our purposes. However, since the intuition behind the coming analysis is significantly more clear for the case of continuous PSD inputs (where we can argue in terms of integrals rather than sums), in the next section we will assume (incorrectly) that injecting an arbitrary PSD into the cavity results in a force function given by integrating the force due to each individual frequency component over the PSD. In the section after we show how this analysis can be adjusted to work for frequency comb inputs which, as we have explained, \emph{can} have their interference effects removed.

\section{Approximation of an arbitrary force function}
\label{sec: Approximation of an arbitrary force function}

Suppose we desire a theoretical force function $F_\textrm{th}(x)$. We wish to find the PSD $p(\delta)$ of an input laser that will cause the total radiation pressure force $F_\textrm{rp}(x)$ experienced by the mirror to be as close to $F_\textrm{th}(x)$ as possible. Note that for convenience we consider the PSD to be a function of the detuning from the central frequency $\omega_0$, and we will assume that the PSD is localised within the particular free spectral range around $\omega_0$. As mentioned in the previous section, we will also assume that no interference effects occur between the different frequency components of the input field. Under these assumptions the force on the end mirror due to the input field $p(\delta)$ is
\begin{align}
	F_\textrm{rp}(x)	&	= \int_{-\infty}^\infty F_x(\delta)p(\delta)\,d\delta			\label{eqn: F_rp definition}\\
				&	= \int_{-\infty}^\infty F_0(\delta+g_0x)p(\delta)\,d\delta 	\nonumber \\
				&	= (F_0\ast p)(-g_0x),\nonumber
\end{align}
where $F_0\ast p$ is the convolution of the force at zero displacement and the PSD. Our goal is therefore to choose $p(\delta)$ so that $F_\textrm{th}(x)\approx F_\textrm{rp}(x)= (F_0\ast p)(-g_0x)$. It will be instructive to write the convolution in the equivalent form
\begin{equation}
	F_\textrm{rp}(x) = (F_0\ast p)(-g_0x) = (F_0/\beta\ast\beta p)(\delta)\big|_{\delta=-g_0x}, \label{eqn: F_rp convolution}
\end{equation}
where $\beta:=\int_{-\infty}^\infty F_0(\delta)\,d\delta$. This suggests that we may view the action of the cavity as transforming the input field into the radiation pressure force function as follows: first a scaling of the input field by $\beta$, then a smoothing by the normalised Lorentzian $F_0/\beta$, and then a change of variable $\delta \to x = -\delta/g_0$. The smoothing, which is analogous to a Gaussian blur, will blur out any features smaller than the linewidth of the cavity, but will preserve larger features. This implies that theoretical force functions with features smaller than the linewidth of the cavity cannot be approximated with this method, since any true force function output from the cavity cannot have such features. On the other hand, if the theoretical force function $F_\textrm{th}$ has only larger features, then it \emph{can} be reliably approximated. This condition translates to requiring that the function be blurred by $F_0/\beta$ without significant effect:
\begin{equation*}
	F_\textrm{th}(-\delta/g_0) \approx (F_0/\beta \ast F_\textrm{th}\big|_{x=-\delta/g_0})(\delta).
\end{equation*}
With the above assumption, we simply need to choose
\begin{equation}
	p(\delta) = F_\textrm{th}(-\delta/g_0)/\beta
	\label{eqn: p setting}
\end{equation}
and then by Eq.~\ref{eqn: F_rp convolution} we have
\begin{align*}
	F_\textrm{rp}(x)	&	= (F_0/\beta\ast F_\textrm{th}\big|_{x=-\delta/g_0})(\delta)\big|_{\delta=-g_0x} \\
				&	\approx F_\textrm{th}(-\delta/g_0)\big|_{\delta=-g_0x} \\
				&	= F_\textrm{th}(x).
\end{align*}
That is, choosing the input field according to Eq.~\ref{eqn: p setting} will cause the force experienced by the mirror to be approximately $F_\textrm{th}$, as required, and the resolution of this approximation is the linewidth of the cavity.

We stress again that this result hinges on the lack of interference effects between the different frequency components of the input field, which is not the case in general. In the next section we show that a similar result holds when the input field is a frequency comb, and we have seen that for these input fields it is possible to perform frequency shifting to eliminate any such relevant interference.

\begin{figure}[]
	\centerline{\includegraphics[width=\columnwidth]{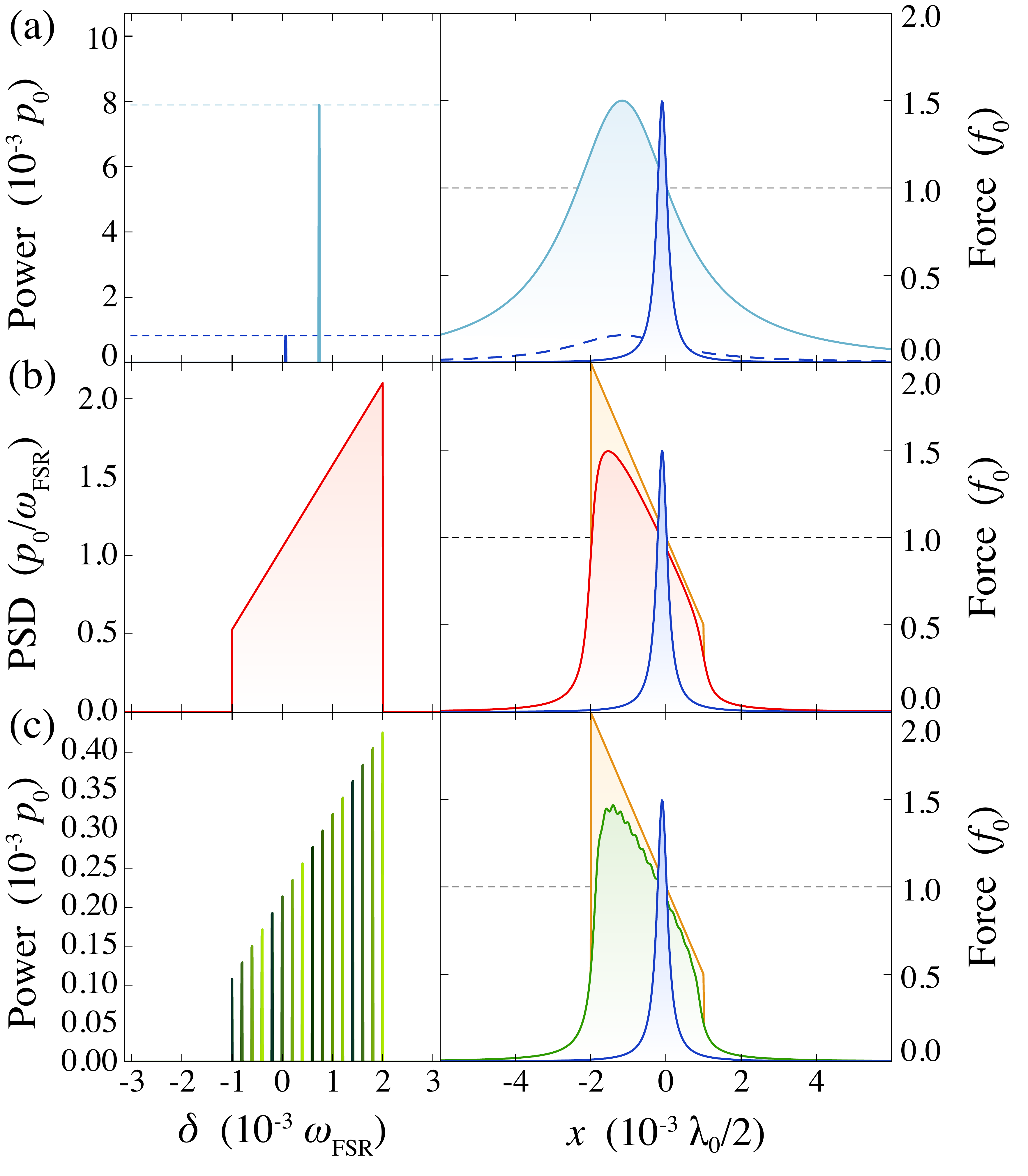}}
	\caption[]{Force functions for different input fields. All quantities are expressed in terms of the wavelength, $\lambda_0$, the weight of the mirror of mass $m$, $f_0 = mg,$
	and the minimum power for levitation, $p_0:=cf_0/2$. We use $\omega_\textrm{FSR} = 2\pi\times\SI{750}{\mega\hertz}$, $\lambda_0=\SI{1064}{\nano\meter}$, $\Delta=0.03\times 10^{-3}\,\omega_\textrm{FSR}$, and $R_\textrm{m}=99.99\%$. The finesse is adjusted by varying the reflectivity of the input mirror, $R_\textrm{i}$. \;\textbf{(a)} Single-frequency inputs and the resulting force functions for cavities with a high finesse of $3000$ (dark) and a low finesse of $300$ (light). To maintain the same potential depth set by the lower trap threshold, a greater input power is required for the low-finesse cavity. If the same input power as the high-finesse case were used, the force would result in the dashed profile, with its maximum noticeably lower than the force $f_0$ required to balance the weight of the mirror. \;\textbf{(b)} Input field with a continuous PSD and the resulting force function in the instance of a high-finesse cavity. The total input power is in this case given by integrating the input field over the frequency domain. The function approximated is an ideal ramp (in yellow), adjusted so that the resulting force function has the same trap threshold as the single-frequency case (shown in blue for comparison). \;\textbf{(c)} Frequency comb as input and the resulting force function when applied to the high-finesse cavity. Each peak represents additional input power, but their total contribution is still lower than the total input power required by the low-finesse cavity in (a). Note that for the ramping input we have plotted the frequencies modulo $\omega_\textrm{FSR}$, with each free spectral range depicted as a different shade of green: using the first peak as reference, the next peak is shifted from it by an amount $\Delta+\omega_\textrm{FSR}$ instead of simply $\Delta$, and so on for the following ones until the total shift from the reference would produce beating effects that would be averaged out within the time scale of the system (in this representative case chosen as $4\omega_\textrm{FSR}$), at which point the process can be repeated on the same free spectral ranges without relevant interference.}
\label{fig: profiles}
\end{figure}

\section{Practical considerations}
\label{sec: Practical considerations}

In this section we show that the use of a frequency comb centred at $\omega_0$ with some given spacing $\Delta$ can be used instead of a continuous PSD input to achieve the required approximation. Applying the appropriate rectangle approximation to Eq.~\ref{eqn: F_rp definition}, we have
\begin{align*}
	F_\textrm{rp}(x)	&	=\int_{-\infty}^\infty F_x(\delta)p(\delta)\,d\delta \\
				&	\approx \sum_{n\in\mathbb{Z}}F_x(n\Delta)p(n\Delta)\Delta\\
				&	=\sum_{n\in \mathbb{Z}}F_x(n\Delta)\frac{F_\textrm{th}(-n\Delta/g_0) \Delta}{\beta},
\end{align*}
where in the last step the PSD was chosen according to Eq.~\ref{eqn: p setting}. The right-hand side corresponds precisely to the force acting on the mirror if the input field is a frequency comb such that the component at frequency $\omega_0+n\Delta$ has power $F_\textrm{th}(-n\Delta/g_0) \Delta/\beta$, assuming interference effects are eliminated. As discussed in section~\ref{sec: The interaction of multiple optical springs}, this elimination can be performed by frequency shifting components of the comb by multiples of the free spectral range.

Therefore a suitable frequency comb yields an approximation of $F_\textrm{rp}$, which is itself an approximation of $F_\textrm{th}$. Note that, unsurprisingly, a smaller spacing $\Delta$ leads to a better approximation.

For many types of force function, the required frequency comb can be generated by modulation of a single mode. Specifically, the components of the comb can be realised as sidebands of the central frequency, with their power determined by the amplitude of the modulation. A combination of phase and amplitude modulation can be used to enforce any asymmetries in the required comb. The width of the comb (the maximum difference between the central frequency and a component of the comb) determines the maximum modulation frequency necessary, and is thus the factor limiting the class of PSDs that can be approximated with a single modulator. Applying a sequence of modulators would allow wider combs to be created, at the expense of simplicity and flexibility. Alternatively, the modes of the frequency comb might be generated by commercial multi-channel laser systems, capable of independently tuning the central frequency of each channel by up to a few tens of $\SI{}{THz}$.

\begin{figure}[]
	\centerline{\includegraphics[width=\columnwidth]{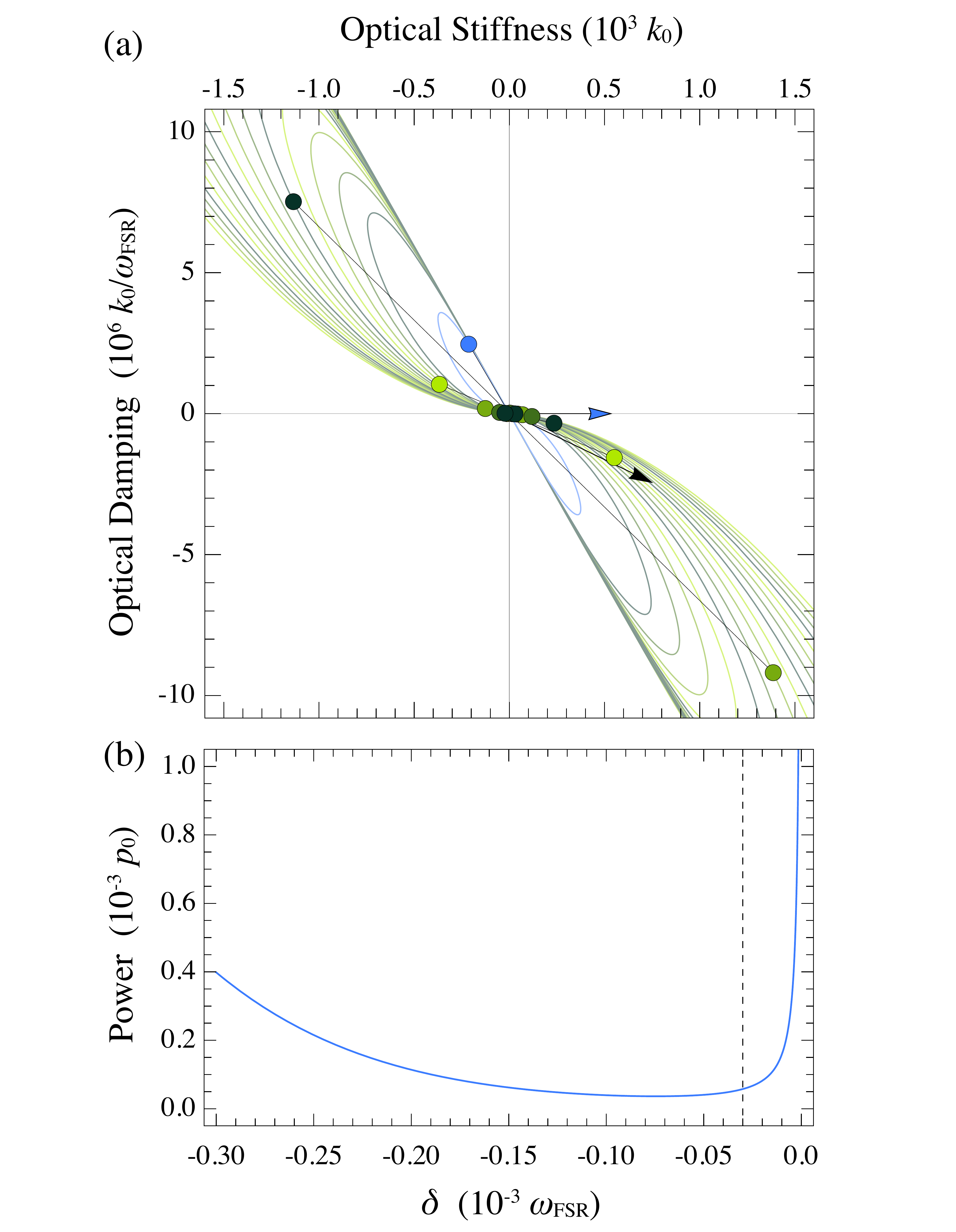}}
	\caption[]{\textbf{(a)} Representation of the optical spring generated by the comb input of Fig.~\ref{fig: profiles} in the complex plane. The scaling is given in units of $k_0 := 2f_0/\lambda_0$. The horizontal axis, $\operatorname{Re}[k_\textrm{os}(\omega)]$, gives the optical stiffness introduced by the cavity; the vertical axis, $-\operatorname{Im}[k_\textrm{os}(\omega)]/\omega$, maps the optical damping due to the imaginary part of Eq.~\ref{eqn: optical spring}. The curves, of different size for different power of the input mode, are parametrized as a function of detuning $\delta\in(-\infty,\infty)$. The individual optical springs of each mode of the comb are indicated by circles on the parametrized curves. As in Fig.~\ref{fig: profiles}c, the curves corresponding to modes in different free spectral ranges are coloured with different shades of green. The combined contribution of the optical springs (black arrow) is equivalent to a single, blue-detuned beam with positive stiffness and negative damping; to avoid parametric instability, an additional, independent cooling beam is considered (blue curve) to cancel the anti-damping effect. The total spring (blue arrow) resulting from the combination of the comb and the cooling beam is purely real and positive. \;\textbf{(b)} Power needed for the cooling beam to cancel the anti-damping effects of the total spring from the comb input, as a function of its detuning. A dashed line indicates the detuning (and power) used in (a). Depending on detuning and power, the total positive optical stiffness resulting may have a different magnitude.}
\label{fig: comb}
\end{figure}

\section{An application: improving the sensitivity of a gravitational sensor}
\label{sec: An application}

In this section we consider a particular optomechanical system, and show how the techniques developed above can improve its sensitivity as a gravitational sensor. We consider a simplified version of the levitation system in \cite{Guccione:2013:PhysRevLett}, where the top mirror of a single vertical cavity is supported by the radiation pressure from the intra-cavity field and constrained to move vertically. Levitation ensures complete mechanical isolation of the top mirror from the environment and allows the detection to be unaffected by external noise. The stability of the levitated system is due to the optical spring effect \cite{Braginsky:1997:PhysLettA,Sheard:2004:PhysRevA,Corbitt:2007:PhysRevLett}. With an input laser blue-detuned from resonance a decrease in cavity length induces a rise in intra-cavity power. The additional radiation pressure will then push the mirror back towards equilibrium. Conversely an increase in cavity length reduces the power and causes the radiation pressure force to drop and the mirror to fall back towards equilibrium. Associated to the positive restoring force there is also negative damping that can foster parametric instability of the oscillator. The instability is easily counterbalanced by the positive dissipative attributes of a second, red-detuned laser~\cite{Corbitt:2007:PhysRevLett} at much lower power. The addition of such a damping laser does not limit the ability to engineer the desired optical spring constant.

For mechanically-clamped oscillators the addition of an optical spring perturbs the original mechanical frequency and stiffness, and the effective spring constants produced can be extremely stiff. The levitated system under investigation is even more extreme, since in the absence of any rigid support the oscillation relies entirely on the optical spring, and the stiffness is fully determined by the slope of the cavity's spectral peak in the blue-detuned region. This property makes the system a particularly illustrative and simple example to consider, but the methods we will apply for its analysis could easily be adapted to far more general systems.

\begin{figure*}[]
	\centerline{\includegraphics[width=\textwidth]{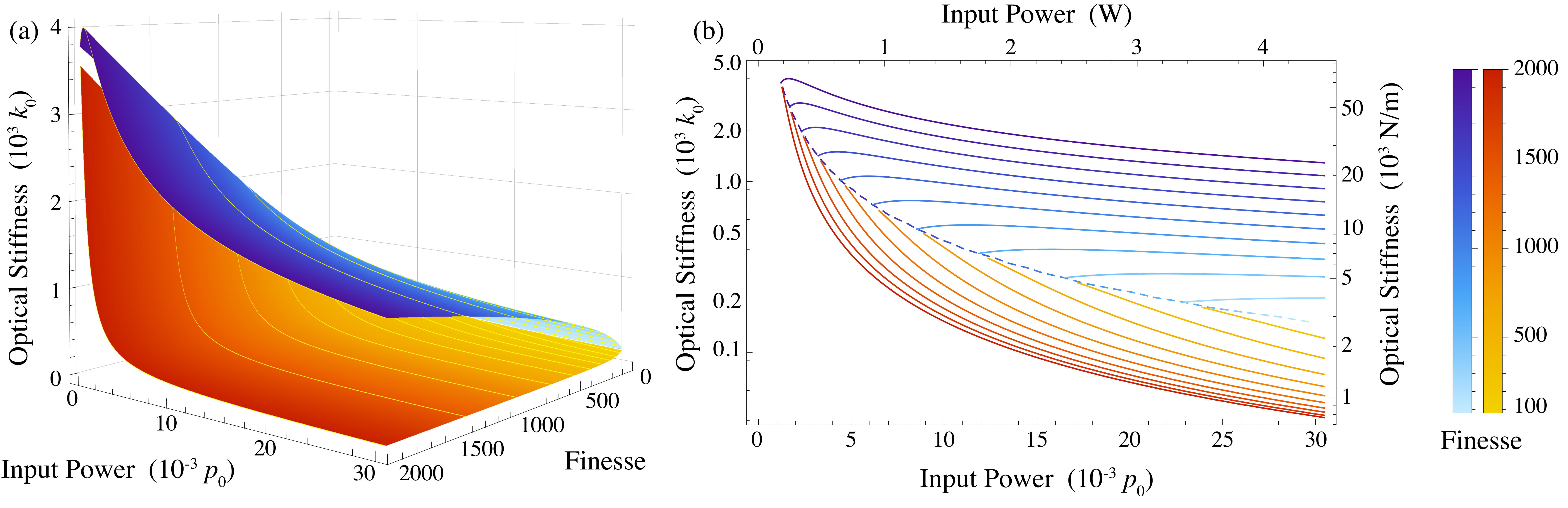}}
	\caption[]{Comparison of the spring constant obtained with a single-mode and a continuous ramp PSD input.
The axis for input power is rescaled in units of the minimum power needed for levitation, $p_0$; the spring constant is expressed in terms of $k_0 = 2f_0/\lambda_0$. Choices for the parameters are: $\lambda_0 = \SI{1064}{\nano\metre}$ for wavelength, $\omega_\textrm{FSR} = 2\pi\times\SI{750}{\mega\hertz}$ for free spectral range, $R_\textrm{m} = 99.99\%$ for the reflectivity of the levitated mirror. The finesse is adjusted by varying the reflectivity of the input mirror, $R_\textrm{i}$. When physical units are given, they correspond to a choice of $m=\SI{1}{\milli\gram}$ for the mirror mass. \textbf{(a)} Full comparison as a function of input power and finesse. The blue surface is the spring constant obtained from a normal cavity. The red surface is the minimum achievable spring constant when a continuous ramp PSD input (red curve in Fig.~\ref{fig: profiles}) is used instead. As finesse decreases, the advantage of approximating an ideal force function is lost because of the lower resolution. \textbf{(b)} Cross sections of the full comparison for different finesses. The starting point of each blue curve corresponds to the minimal stability condition, that the maximum of the force function should be exactly equal to $1.5 f_0$, and the locus of such points is shown as a dashed line; further increasing the power shifts the balancing point $f_0$ along the profile of the Lorentzian force function, yielding different spring constants. Each point on a red curve corresponds to the lowest spring constant achievable given the minimal stability condition, and now the effect of having more input power available is to allow for a wider ramp, which reduces the slope of the force function while keeping the maximum value fixed at $1.5 f_0$.}
\label{fig: tradeoff}
\end{figure*}

The mirror's position can be probed very precisely via the reflected (or transmitted) optical field \cite{Aspelmeyer:2014:RevModPhys}, and changes in weight inducing a variation of the mean position can be monitored to deduce relative variations of the gravitational constant, $g$. Sensitivity can be increased if the same weight fluctuation can induce a larger change in position, and this is attained by having a smaller (softer) optical spring constant. We note that decreasing the spring constant makes the system more susceptible to radiation pressure noise introduced by quantum amplitude fluctuations, but feedback and sensing methods have been suggested~\cite{Zach-Korth:2013:PhysRevA} to push the force measurement sensitivity beyond this limitation. Furthermore, in the course of minimising the spring constant we must ensure that conditions for stable levitation still hold, since a too-high reduction in stiffness would cause the mirror to fall out of the trapping region even with slight fluctuations. Thus we consider only methods to reduce the spring constant which allow the trap threshold, defined to be the maximum value of the force function or equivalently the maximum weight supported, to remain constant.

The simplest way to lower the optical spring constant is to reduce the finesse of the cavity, but unless the input power is increased this causes a reduction in the potential depth defined by the trap threshold. Indeed, integrating both sides of Eq.~\ref{eqn: F_rp definition} and switching the order of integration shows that the total input power to the cavity is proportional to the integral of the force function\footnote{The ``constant'' of proportionality here is $g_0/\beta$, which depends on the cavity finesse, but it is readily verified that if the reflectivity of the moveable mirror is held constant at a value close to $1$ then $\beta$ is essentially constant when the finesse is any more than around $10$.}, and it is clear geometrically that regardless of the specific shape of the force function, to maintain the depth of the trap while reducing the spring constant the integral of the force function (and therefore the input power) must increase. Put another way, for a given trap threshold there is a trade-off between reducing the spring constant to improve the sensitivity, and keeping the input power low due to availability or to avoid damaging the optics. There is one more factor to consider, which is the effect of cavity finesse on the precision of the position measurement. The simplest method for achieving a high-precision readout of the mirror position is to measure the phase shift in the field reflected or transmitted from the cavity, but the sensitivity of this approach scales with cavity finesse, since a higher finesse causes photons to accumulate a larger phase shift (for a given mirror displacement) before leaving the cavity. Thus we see that lowering the finesse to reduce the spring constant could potentially have a negative effect on the overall sensitivity of the system. As we will see, the method presented above allows us to both overcome this problem, and optimise the aforementioned trade-off between stiffness and input power beyond what is achievable by simply changing the cavity finesse.

In Fig.~\ref{fig: profiles} we present a comparison between the force functions we would expect from single-mode inputs to high- and low-finesse cavities, an ideal force function, and two approximations of this ideal function. The ideal force function considered is tailored to the task at hand, aiming to achieve a lower spring constant for the same trap threshold, consisting of a ramp with a gentle slope around the equilibrium position that drops off immediately outside the linear region in order to minimise input power. Note that this ideal force function extends further in the ``blue-detuning'' region to ensure that the approximating functions have a trap threshold equal to that of the single-mode functions. The two approximating functions are obtained with an ideal, but unattainable, continuous PSD input and with a plausible frequency comb input applied to high-finesse cavities. Comparing the high- and low-finesse curves of Fig.~\ref{fig: profiles}a we see that a lower finesse leads to a softer spring constant, but the force function has a larger integral and generating it requires more input power. Comparing these with the multi-mode inputs of Fig.~\ref{fig: profiles}b and~\ref{fig: profiles}c, we see that an appropriate choice of input field can reduce the spring constant significantly below that of even the low-finesse function without the same power requirements. Splitting the available power into multiple modes to make the comb wider leads to another advantage, which is that the mirror becomes more robust to larger displacements by being subject to a similar stiffness over a larger range. In Fig.~\ref{fig: comb}a it can be seen that for a specific mirror position the total spring is mostly generated by only a few peaks of the comb; the other peaks, whose contributions in this configuration are gathered around the origin, would become involved when the mirror is displaced, at which point they would swap roles with the peaks resonating in the original configuration and thus maintain the stiffness constant.

As in the case of a single mode input, a multi-mode input can produce a force acting on the mirror which is restoring but also anti-damping if the effective contributions are in the blue-detuned regime. This can be seen in Fig.~\ref{fig: comb}a, where the resultant optical spring has positive stiffness but negative damping. A solution to this problem is provided by the introduction of an additional red-detuned beam, specifically tuned to cancel any anti-damping effects without significantly affecting the restoring force tailored by the original choice of input field.

The comparison between single- and multi-mode inputs for different finesses can be appreciated more quantitatively in Fig.~\ref{fig: tradeoff}, where the spring constant value is shown as a function of input power and finesse when the input field is either single-mode or the ramping, continuous PSD. In this example we use an ideal, non-interfering, continuous PSD instead of a more physical comb input, as the former allows an easier mathematical formulation while the results obtained would be comparable in the limit of a very fine comb structure. For the single-mode input we impose the stability constraint that the trap threshold be at least $1.5$ times larger than the weight of the mirror. If either finesse or power are too low to achieve this condition then no spring value can be considered. Given a specific finesse, a change increasing the power above the minimum requirement drives the stability condition to a different point of the familiar Lorentzian profile of the cavity, and lower spring values are obtained when the input power is high enough to have equilibrium closer to the base. A higher finesse generally implies a higher spring constant, as the gradient of the Lorentzian is steeper. For a ramp input we impose the same stability condition as before, but due to the extra degree of freedom afforded by the choice of ramp function, and the goal of minimising the achievable spring constants, it is optimal to always use trap thresholds of exactly $1.5$ times the mirror weight (since any power used to further increase the trap threshold would be better utilised for decreasing the spring constant). In this case we see that for any given power and finesse, significantly softer spring constants are obtained when compared to the single-mode case. In fact, for a given power, even if different finesses are used for each method (provided they are high enough to satisfy the stability conditions) it is always advantageous to use the ramp input. Moreover, unlike with a single-mode input, increasing the finesse improves the quality of the approximation and leads to a greater advantage.

Thus we see that even for the very simple class of force functions approximated in this example the extra flexibility made possible by our method allows for significant enhancements in performance.

\section{Conclusion}
In conclusion, we have presented a simple analysis demonstrating that, by judicious choice of input field, the potential experienced by a moveable end mirror of a linear cavity can be can be tailored to perform more effectively for the task of interest. This tailoring is possible whenever the desired potential possesses features no finer than the cavity linewidth, and the cavity length varies by amounts small relative to the wavelength of the light. Given the practical difficulty in obtaining specific PSDs, we have further shown that the use of a frequency comb is a viable alternative in terms of the quality of the approximation, and that for many force functions the required combs can be generated by simply applying phase and amplitude modulation to a single mode. Finally, we have given a concrete example showing a possible application of this technique. Given its simplicity and generality, the method presented could conceivably be used in a wide variety of optomechanical systems as a simple way to improve performance.

\section{Acknowledgements}
This research is supported by the Australian Research Council Centre of Excellences CE110001027, the Discovery Project DP150101035. PKL is supported by the ARC Laureate Fellowship FL150100019, BCB by the ARC future fellowship FT100100048. This paper has the LIGO document number LIGO-P1300068.

\appendix
\section{Numerical estimates for the effects of interference}
To estimate the effects that the beating of several input fields has on the mechanical system we consider a simplified system, where two input fields are detuned by some amount $\omega_\textrm{det}$ relative to one another. In the worst case scenario, the two fields have identical strength and the total power is modulated between perfect destructive and constructive interference. We use the assumption that each input, independently, produces the minimum intra-cavity power $p_0=mgc/2$ necessary for levitation of the system discussed in section~\ref{sec: An application}. The beating of the two inputs results in the time-dependent power
\begin{equation*}
	P(t)=2p_0\cos^2{(\omega_\textrm{det}t)},
\end{equation*}
and as a consequence the radiation pressure force also becomes a function of time:
\begin{equation*}
	F(t)\simeq\frac{2P(t)}{c}=2mg\cos^2{(\omega_\textrm{det}t)}.
\end{equation*}
The dynamics of the mechanical system are described by a differential equation for $x$ which includes the restoring and the radiation pressure force terms,
\begin{equation*}
	m\ddot{x}(t)=-mg-m\omega_\textrm{m}^2x(t)+2mg\cos^2{(\omega_\textrm{det}t)}.
\end{equation*}
In the limit where the time scale of the beating is much faster than that of the mechanical system, $\omega_\textrm{det}\gg\omega_\textrm{m}$, the solution to the equation is
\begin{equation*}
	x(t)=-\frac{g}{4\omega_\textrm{det}^2}\cos{(2\omega_\textrm{det}t)}.
\end{equation*}
We see that these oscillations can be made arbitrarily small by choosing a large enough detuning $\omega_\textrm{det}$ between the input fields. For example, if we choose $\omega_\textrm{det}$ equal to the free spectral range considered in section~\ref{sec: An application}, $\omega_\textrm{FSR}=2\pi\times\SI{750}{\mega\hertz}$, the off-resonance oscillations induced by the beating have amplitude on the order of $\SI{E-9}{\angstrom}$. This is even smaller than the amplitude of the position's zero-point quantum fluctuations, $x_\textrm{ZPF}=\sqrt{\hbar/(2m\omega_\textrm{m})}$, which is on the order of $\SI{E-8}{\angstrom}$ for a mirror of mass $m=\SI{1}{\milli\gram}$ and frequency of the oscillations $\omega_\textrm{m}=2\pi\times\SI{1}{\mega\hertz}$.

Being able to minimize the effect on the mechanics might not be enough, however, as the oscillation driven at $2\omega_\textrm{det}$ may also coherently interact with the cavity modes and resonantly alter the response of the optical resonator. The induced oscillations of the mirror function as a source of frequency modulation for the field, creating sidebands resonating at frequencies $2\omega_\textrm{det}$ away. It is important to check that these sidebands have negligible effect on the system. Assuming the cavity to be on resonance when the mirror is at the centre of the oscillation, the equation governing the dynamics of the cavity field $\alpha$ is, in terms of the input field $\alpha_\textrm{in}$,
\begin{equation*}
	\dot{\alpha}(t)=\left[-\gamma/2+ig_0x(t)\right]\alpha(t)+\sqrt{\gamma}\alpha_\textrm{in}(t).
\end{equation*}
In the Fourier domain, where position acquires the form of a double delta function $x(\omega)=-\pi g/(4\omega_\textrm{det}^2)\left[\delta(\omega-2\omega_\textrm{det})+\delta(\omega+2\omega_\textrm{det})\right]$, the equation for the cavity field becomes
\begin{widetext}
\begin{equation*}
	(\gamma/2-i\omega)\alpha(\omega)-i\frac{\pi g}{2\lambda_0\omega_\textrm{det}}\left[\alpha(\omega-2\omega_\textrm{det})+\alpha(\omega+2\omega_\textrm{det})\right]=\sqrt{\gamma}\alpha_\textrm{in}(\omega).
\end{equation*}
\end{widetext}
The sidebands arising from the frequency modulation have, therefore, an amplitude on the scale of $\pi g/(\gamma\lambda_0\omega_\textrm{det})$ relative to their carriers. If we choose once more $\omega_\textrm{det}=\omega_\textrm{FSR}$ and a finesse of $1000$, this is equivalent to only about $1$ part per billion.

%\bibliographystyle{plainbib}
%\bibliography{paper.bib}

\end{document}